\documentclass{jfm}
\usepackage{amssymb,latexsym,bm,amsmath,epsfig}

\usepackage[OT1]{fontenc}
\usepackage[applemac]{inputenc}

\catcode`\ú=\active \gdefú{\setbox0=\hbox{0}\hbox to\wd0{}}
\def\Rey{\mbox{\it Re}}   

\ifx\pdfoutput\undefined
\usepackage{graphicx}
\graphicspath{{figuresEPS/}}
\else
\usepackage[pdftex]{graphicx}
\usepackage{epstopdf}
\graphicspath{{figuresEPS/}}
\fi
\title[Shear effects on passive scalar spectra]{Shear effects on
passive scalar spectra}

\author[A. Celani, M. Cencini, M. Vergassola, E. Villermaux and D. Vincenzi]%
{A.\ns C\ls E\ls L\ls A\ls N\ls I$^{1,2}$%
,\ns
M.\ns C\ls E\ls N\ls C\ls I\ls N\ls I$^3$%
,\ns
M.\ns V\ls E\ls R\ls G\ls A\ls S\ls S\ls O\ls L\ls A$^4$\ls,
\ns
E.\ns V\ls I\ls L\ls L\ls E\ls R\ls M\ls A\ls U\ls X$^5$
,\ns 
\and \ls D.\ns V\ls I\ls N\ls C\ls E\ls N\ls Z\ls I$^{2,6}$ }

\affiliation{ $^1$ CNRS, INLN, 1361 Route des Lucioles, F-06560
Valbonne, France.\\
[\affilskip] $^2$ CNRS UMR 6202, Observatoire de la C\^ote 
d'Azur, B.P. 4229, 06304 Nice Cedex 4, France.\\
[\affilskip] $^3$ Center for Statistical Mechanics
and Complexity, INFM Roma 1 and Dipartimento di Fisica,
Universit\`a di Roma ``La Sapienza", Piazzale Aldo Moro, 2, I-00185
Roma, Italy. \\
[\affilskip] $^4$ CNRS, URA 2171, Institut Pasteur, 28, rue du Dr. Roux, 75724 Paris 
Cedex 15, France.\\
[\affilskip] $^5$ Univ. Aix Marseille 1, IRPHE, 49, rue
Frederic Joliot Curie, F-13384 Marseille 13, France.\\
[\affilskip] $^6$ Dep. de Math\'ematiques, Univ. de Nice-Sophia
Antipolis, Parc Valrose, 06108 Nice, France.
}

\pubyear{????}
\volume{???}
\pagerange{??--???}
\begin{document}

\maketitle

\begin{abstract}
The effects of a large-scale shear on the energy spectrum of a
passively advected scalar field are investigated. The shear is
superimposed on a turbulent isotropic flow, yielding an
Obukhov-Corrsin $k^{-5/3}$ scalar spectrum at small scales. Shear effects
appear at large scales, where a different, anisotropic behavior is
observed. The scalar spectrum is shown to behave as $k^{-4/3}$ for a shear fixed in 
intensity and direction. For other types of shear characteristics, 
the slope is generally intermediate between the $-5/3$ Obukhov-Corrsin's and
the $-1$ Batchelor's values.  The
physical mechanisms at the origin of this behaviour are illustrated in
terms of the motion of Lagrangian particles. They provide an
explanation to the scalar spectra shallow and
dependent on the experimental conditions observed in shear flows at
moderate Reynolds numbers.
\end{abstract}
\section{Introduction}
Two major regimes are known for a passive scalar field transported by
a turbulent flow. The first, known as inertial-convective, refers to
the range of scales where the direct effects of the fluid viscosity
and the scalar diffusivity are both negligible. The predicted behaviour
of the spectrum $E_{\theta}(k)$ of the scalar field $\theta$ is
(\cite{Obukhov49,Corrsin51}):
\begin{equation}
E_{\theta}(k)= C\epsilon^{-1/3} \epsilon_{\theta}\,\, k^{-5/3}\,.
\label{eq:1}
\end{equation}
Here, $k$ denotes the wavenumbers, $C$ is an adimensional constant,
$\epsilon$ and $\epsilon_\theta$ are the average dissipation rates of
the kinetic energy and the scalar variance, respectively.  The second
regime, dubbed viscous-convective, is at scales smaller than
the former and holds for weakly diffusive (alternatively high Schmidt, or Prandtl number) 
scalars. The
velocity field is now smoothed out by the molecular viscosity, linearizing 
the dependence of the velocity increments on the separation 
between the two points where the increment is
taken. That yields the Batchelor behaviour (\cite{Batchelor59}):
\begin{equation}
E_{\theta}(k) \propto k^{-1}\,.
\label{eq:1b}
\end{equation}

Scalar spectra have been measured in a number of laboratory
experiments. None of these pure power laws with the above anticipated 
exponents are routinely observed. The existence of a $k^{-1}$ in the viscous-convective 
subrange has been questioned based on high Reynolds and Schmidt numbers experiments 
(\cite{Miller96}, \cite{Warhaft00}, \cite{Yeung02}). As for larger scales, 
scalar spectra do behave 
compatibly with a power law,
but their slope is often found to deviate from the
Obukhov-Corrsin $5/3$ value. Deviations are particularly important in
turbulent shear flows, where the observations indicate that the
spectra strongly depend on the Reynolds number \Rey$\;$
(\cite{Mestayer82,Sreenivasan91,Sreenivasan96,Miller96,Warhaft00}) and on the
detailed set-up of the advecting turbulent flow. The prediction
(\ref{eq:1}) is in particular recovered for very large Reynolds
numbers only ($\Rey_{\lambda} > 2000$).  Deviations are observed
also when the velocity field displays a convincing $k^{-5/3}$ spectrum
(see, e.g., \cite{Sreenivasan96,Villermaux01}).

Our aim here is to investigate the physical origin of the behaviour
experimentally observed in shear flows, briefly recalled in ~\S\,2. 
We specifically consider a model where the
velocity is decomposed into a linear shear, superimposed onto a
turbulent flow supposed to obey Kolmogorov's scaling. 
As in thermal convection (see, e.g., \cite{Siggia94}), 
the presence of the shear introduces 
a new length-scale $r_c$. The scalar spectrum crosses from the classical 
behavior (\ref{eq:1}) at scales $r\ll r_c$ to a different anisotropic behavior for 
$r\gg r_c$. As discussed in ~\S\,3, Lagrangian 
dimensional arguments allow to predict 
the scaling of the scalar spectrum in the new shear-regime for a variety of shear 
properties. A $-4/3$ scaling is, in particular, 
found for a shear fixed in intensity and direction.  
A test of the Lagrangian dimensional arguments is provided 
in ~\S\,4, where we consider 
a simpler turbulent flow with a short correlation
time. The assumption is unphysical and 
the very same 
Lagrangian arguments of ~\S\,3 yield quantitatively different 
predictions; 
the point is however that these predictions can be 
stringently tested against the exact solution for the scalar spectrum  
which can be obtained in this special case. 
The final Section is
devoted to discussions and conclusions. The Appendix is reserved to
technical material used in ~\S\,4.

\section{Scalar spectra in experiments}

Turbulent flows might be divided into two broad categories: shear flows
(jets, wakes, and boundary layers) and flows where the effects of
large-scale shears are tamed (e.g., grid-generated turbulence).  The
latter are the laboratory flows providing the closest physical
realization of the homogeneous and isotropic turbulence which most of
the theoretical fluid dynamics deals with.

Various possible ways to inject scalar fields in turbulent flows have
been employed.  In grid turbulence, temperature fluctuations are generated
by heating either the grid itself or small wires located thereafter.
In shear flows, one possibility is to weakly heat the jet or the wall
of the boundary layer. Alternatively, a colorant dye can be injected
past the flow source, at a scale possibly different from that of 
the stirring source.  Temperature time series are taken by a 
thermometer in a fixed location and, as customary in most fluid dynamics experiments, time 
distances are converted to space distances by Taylor hypothesis.  Scalar spectra or, equivalently, 
second-order structure functions, are thus measured. The use of fluorescent dyes offers the 
possibility to extract cuts --usually two-dimensional-- 
of the scalar field from the intensity field.  Simultaneous velocity
measurements are usually taken by anemometers.  The reader interested
in a detailed discussion of the experimental techniques is referred to
\cite{Sreenivasan96,Warhaft00,Villermaux01}.

The observations gathered in the literature might be 
summarized as follows. In grid-generated turbulence, 
the prediction (\ref{eq:1}) is well
verified. That holds also at moderate $\Rey_{\lambda}$, when there is
just a tiny range of scales where the spectrum of the velocity field
displays a clean $-5/3$ slope.  Some results obtained heating directly
the grid used to generate the turbulence have led to some criticisms
(\cite{Warhaft00}) on account of possible correlations between the
temperature and the velocity.  It seems,
nevertheless, a well-established fact (see
\cite{Jayesh94,Mydlarski98,Warhaft00}) that the slope of the scalar spectrum 
increases with $\Rey_{\lambda}$ from values $\simeq 1.5$ to $5/3$,
reached at Reynolds numbers $\Rey_{\lambda}\geq 200$. Small
deviations observed at lower $\Rey_{\lambda}$ are confidently ascribed
to finite \Rey$\;$-number effects.

The situation in turbulent shear flows is more intricate.  Experiments
with different $\Rey_{\lambda}$'s but similar non-dimensional shear
rates have been collected and compared by \cite{Sreenivasan96}.  The
main observation is that the slopes of the passive scalar spectra
display a strong dependence on the turbulent intensity. Their slope is
approximately $1.3$ at low $\Rey_{\lambda}\,$ and attains the
predicted $5/3$ value at $\Rey_{\lambda}\geq 2000$ only. Strong
anisotropic effects are also observed for the velocity energy
spectrum. Indeed, its longitudinal component reaches the expected
$-5/3$ behaviour at moderate $\Rey_{\lambda}\, (\approx
50)$. Conversely, its transverse component behaves more similarly to
the scalar field, with a slope continuously increasing with
$\Rey_{\lambda}$ and saturating to $5/3$ only for $\Rey_{\lambda} \geq
3000$. The experiments by \cite{Villermaux01} report even more severe
deviations from (\ref{eq:1}), with a slope close to the Batchelor
unit value, in spite of the $-5/3$ spectrum observed for the velocity
field. The scalar injection scale was, 
although still lying in the inertial range of scales, 
smaller than the turbulence forcing scale in that case.

\section{Shear effects on Lagrangian dynamics}

The experimental observations presented in the previous Section
naturally lead to surmise that the deviations observed in the scalar spectra have, 
among other possibles causes, their origin in the presence of a sustained large-scale shear. 
The interplay between diffusion and shear is a classical problem first 
examined by \cite{Leveque28} for heat transport across boundary layers. 
Qualitatively, the flow stretches the distances between particles along the shear and sharpens 
the scalar gradient perpendicular to it, therefore altering the usual diffusion law.

In order to
quantitatively assess the impact of shear on scalar spectra, we
introduce the following simple model for the velocity field ${\bm
v}({\bm r},t)$:
\begin{equation}
{\bm v}({\bm r},t)=\sigma y\, \hat{{\bm x}}+{\bm u}({\bm r},t)\,.
\label{eq:3}
\end{equation}
The flow is thus the superposition of an average linear shear of
intensity $\sigma$ directed along the $x-$direction and a turbulent
fluctuating field ${\bm u}$, supposed to obey 
Kolmogorov's scaling, for the sake of simplicity. 
The spectrum of the velocity field obeys in particular the classical $k^{-5/3}$ law. 

A passive scalar field, $\theta({\bm r},t)$, transported by the above flow is governed by 
the standard ad\-vection-diffusion equation:
\begin{equation}
\partial_t \theta+\sigma y \partial_x \theta +{\bm u} \cdot {\bm
\nabla} \theta=\kappa\Delta \theta+f\,,
\label{eq:6}
\end{equation}
where $\kappa$ is the molecular diffusivity and $f$ is an external
source of scalar fluctuations. The source is needed to maintain the
system in a statistically steady state and we shall suppose that the
forcing has a characteristic length $L_f$.

The transport equation (\ref{eq:6}) is equivalently recast in terms of
the Lagrangian trajectories ${\bm \rho}(t)$ of tracer particles. They
obey the stochastic differential equation
\begin{equation}
d{\bm \rho}={\bm v}({\bm \rho},t)\,dt+\sqrt{2\kappa}\,d{\bm W}\,,
\label{eq:br}
\end{equation}
where ${\bm W}$ is an isotropic Brownian motion and ${\bm v}$ is given
by (\ref{eq:3}). The equation for the scalar field along the
trajectories is the ordinary differential equation $d\theta/dt=f$,
easily integrated. The scalar second-order correlation function
of the scalar field is then expressed as (see, e.g.,
\cite{Falkovich01}):
\begin{equation}
C_2({\bm r},t)=\langle\theta({\bm r},t)\theta({\bm 0},t)\rangle=
\int_{-\infty}^{t} \!\!\!{\rm d}s \int {\rm d}{\bm R}\;
p({\bm R},s|{\bm r},t)\, \chi({\bm R})\,.
\label{eq:10}
\end{equation}
Here, 
$\bm R$ is the separation
between two tracer particles and $p({\bm R},s|{\bm r},t)$ is the
propagator, i.e. the probability, averaged over the realizations of the
velocity ensemble and of the Brownian noise, that a pair of particles
is found separated by ${\bm R}$ at time $s$, conditional to its
separation ${\bm r}$ at time $t$. The function $\chi({\bm R})$
having its support at scales $R\leq L_f$, the integral (\ref{eq:10}) is
roughly equal to $\chi(0)T(r,L_f)$, where $T(r,L_f)$ is the residence 
time at distances $\leq L_f$ for a pair of particles initially separated by
a distance $r$. Analogously, the second-order structure function
$S_2(r)=2\left[C_2(0)-C_2(r)\right]\propto T(0,L_f)-T(r,L_f)$.  The
incompressibility of the flow ensures that the scaling of the latter quantity 
can be estimated as the time for two particles, initially coinciding, to
reach a separation $r$. 

The problem of determining the scaling
behaviour of scalar spectra is thus reduced to the study of the
evolution of the separation between a pair of particles.

For the case without shear, $\sigma=0$, classical dimensional
arguments give the Richardson law: $\langle R^2(t) \rangle\sim t^{3}$,
which translates into a $2/3$ scaling exponent for $S_2(r)$
and the ensuing Obukhov-Corrsin $-5/3$ slope for the scalar spectrum.

Let us now consider the case with shear, $\sigma \neq 0$. The
components of the separation between two fluid particles obey the
equations of motion:
\begin{eqnarray}
\dot{R}_x&=& \sigma R_y + \delta u_x({\bm R},t) \label{eq:12a}\,,\\
\dot{{\bm R}}_{\perp}\!&=&  \delta {\bm u}_{\perp}({\bm R},t)\,, 
\label{eq:12b}
\end{eqnarray}
where the dot indicates the time-derivative, $\delta {\bm u}$ is the
velocity difference between the two particles and $\dot{{\bm R}}_{\perp}$
indicates the components of the velocity orthogonal to the shear.  At
small enough times, the turbulent component, which scales with
exponent $1/3$, dominates over the linear shear and Richardson's
behaviour holds. The crossover to a different anisotropic behaviour
occurs at $r_c \simeq \sigma^{-3/2}$. 
Those are the separations where the shear and the
turbulent components become comparable. The time required for
two particles initially coinciding to reach those scales behaves as 
$t_c\propto \sigma^{-1}$.
For $t\gg t_c$,
the form of the Lagrangian equations naturally suggests that 
the parallel and the transverse components scale differently with time:
\begin{equation}
\langle R_x^2\rangle \approx \langle R^2\rangle \propto
t^{2\alpha}, \qquad {\rm and} \qquad \langle R_{\perp}^2\rangle \sim t^{2\beta}\,.
\label{eq:nuova}
\end{equation}
Inserting this ansatz into (\ref{eq:12a}) and (\ref{eq:12b}), 
neglecting the $\delta u_x$ term in (\ref{eq:12a}) 
and using the Kolmogorov's scaling $\delta {\bm u}_{\perp}({\bm R},t)\propto R^{1/3} \simeq 
t^{\alpha/3}$,  we obtain the relations: 
$\alpha-1=\beta$ and $\beta-1=\alpha/3$.
We thus end up with the predictions $\alpha=3,\;\beta=2$ 
for the Lagrangian separations, that give 
\begin{equation}
C_2(r)\propto r^{1/3};\qquad E(k)\propto k^{-4/3}\,,
\label{spettro}
\end{equation}
for the correlations of the scalar field, by using the arguments following (\ref{eq:10}).

It is worth emphasizing that the scalar statistics
is sensitive to the shear characteristics, as it should 
be expected for any passive transport. 
For example, if the direction of the shear were to
rotate rapidly, isotropy would be recovered and the stretching of the 
separations among the
particles would be dominated by the linear shear component, leading to
an exponential-in-time separation and a $k^{-1}$ behaviour. 
The physics is the same as in the Batchelor regime, but the 
behaviour now holds at {\em large scales}. 
The rapidity of the rotation should be gauged with respect to the 
Lagrangian turn-over time: if $\tau_{\phi}$ denotes the typical time of 
rotation, the fixed limit (\ref{eq:3}) holds for 
scales $r\ll \tau_{\phi}^{3/2}$ and 
the rapidly rotating asymptotics in the opposite limit. As for the 
fluctuations of the shear intensity, let us similarly denote by  
$\tau_{\sigma}$ its correlation time. The intensity  
is effectively fixed in time, as in (\ref{eq:3}), 
for scales $r\ll \tau_{\sigma}^{3/2}$.
In the opposite limit, the intensity 
might be taken as a random process with a short
correlation time $\langle \sigma(t)\sigma(t')\rangle=\sigma^2\delta(t-t')$. 
Inserting the ansatz (\ref{eq:nuova}) into (\ref{eq:12a}) 
leads now to the relation $\alpha-1=-1/2+\beta$, giving a 
slope $-13/9$ for the scalar spectrum. The geometry 
of the shear and the ratio between
its time-scale and the turbulent turn-over times are thus seen to affect
the properties of a passively transported
field. The point of interest to interpret the 
experimental data is that the exponents of the spectra are 
in all cases smaller than $5/3$, implying that the presence of a shear 
generally induces a crossover to 
a shallower spectrum at the large scales. 

\section{A solvable case}

The aim of this Section is to test the dimensional arguments presented in the 
previous section by investigating the simpler case
of a turbulent flow ${\bm u}$ belonging to the Kraichnan
ensemble (\cite{Kraichnan68}). 
The {\it rationale} is as follows. The short correlation time of the 
velocity is unphysical 
and the Lagrangian predictions (\ref{spettro}) are modified 
thereby (the scaling $-4/3$ of 
the scalar spectrum becoming, for example, $-11/9$). 
Yet, the point is that one can compare those predictions {\it versus} 
the solution of the exact equation for the scalar spectrum. The 
advantage over a direct numerical 
simulation of the more realistic flows in Section~3 is that the range 
of scales available here
is much more extended and the comparison 
is therefore quantitatively more stringent, albeit less direct. 

The field ${\bm u}$ in (\ref{eq:3}) is taken now as an
incompressible, statistically isotropic and homogeneous Gaussian field
of zero mean and correlation function:
\begin{equation}
\langle [u_i ({\bm r},t)-u_i ({\bm 0},t)][u_j ({\bm r},0)-u_j ({\bm
0},0)]\rangle=\delta(t)
\left\{D \,r^{\xi}\left[(1+\xi)\delta_{ij}-\xi 
\frac{r_{i}r_{j}}{r^2}\right]\right\}\,.
\label{eq:4}
\end{equation}
Here, we specialize to the two-dimensional case for the sake of simplicity:
this choice bears no qualitative consequences on the results.
The parameter $D$ measures the turbulent intensity and the velocity
field is assumed to be scale-invariant with roughness exponent $\xi\in
[0,2]$. The presence of an ultraviolet viscous cutoff mimicking the
Kolmogorov scale might be considered, but it will not be needed here.

Using standard methods of Gaussian 
integration by parts (see \cite{Falkovich01}), 
we can re-adapt the Lagrangian arguments of the previous section
to the case of a Kraichnan flow. The Lagrangian separation 
law for the isotropic case 
is: $\langle R^2(t)\rangle \propto t^{2/(2-\xi)}$. 
Note that the Richardson scaling is obtained 
for $\xi=4/3$, differing from the $2/3$ 
of real turbulent flows due to 
the $\delta$-correlation in time of the velocity field. For the anisotropic case 
of a shear fixed in intensity and direction, 
the relations between the exponents $\alpha$ and $\beta$ 
stemming from (\ref{eq:12a}) and (\ref{eq:12b}) read now:
$\alpha-1=\beta$  and $2\beta-1=2\alpha\xi$. 
Solving these relations 
gives the following behaviors of the Lagrangian separations and the 
scalar spectra:
\begin{equation}
\langle R_x^2\rangle \approx \langle R^2\rangle \propto 
t^{6/(2-\xi)};\quad \langle R_y^2\rangle \sim t^{2(1+\xi)/(2-\xi)};\quad
E(k)\propto k^{-(5-\xi)/3}\,.
\label{eq:xx}
\end{equation}
For $\xi=4/3$, (\ref{eq:xx}) gives a $-11/9$ exponent for the 
scaling of the scalar spectrum.

The advantage of Kraichnan's flows is that the behavior of 
the second-order correlation
function $C_2(r)$ at the stationary state 
can be determined exactly and used to test the previous
dimensional arguments. In the shear-free case, $\sigma=0$, 
the problem is isotropic and the
expression of $C_2(r)$ is well-known analytically. The inertial-range scaling
behaviour for the second-order structure function 
$S_2(r)=\frac{\chi(0)}{(2-\xi)D}r^{2-\xi}$,
where $\chi(0)$ is the scalar variance injection rate.  The exponent
$2-\xi$ is in agreement with the Lagrangian dimensional estimate. 

In the presence of shear, the equation for the second-order
correlation function is still closed, but the problem is not isotropic
anymore. Following the same procedure as in the shear-free case
(\cite{Kraichnan68}), one obtains:
\begin{eqnarray}
\partial_t C_2&+&
\sigma(r \sin\phi \cos\phi\,
\partial_r-\sin^2\!\phi\,\partial_\phi)C_2
= \frac{1}{r} \partial_r (Dr^{1+\xi}+ 2\kappa r)\partial_r C_2  \nonumber\\
&+&\frac{1}{r^2}[(1+\xi)Dr^\xi+2\kappa]\partial^2_\phi C_2
+\chi(r),
\label{eq:7}
\end{eqnarray}
where $\bm r=(r,\phi)$ denotes the relative distance between the two points. The
forcing is taken Gaussian, statistically 
homogeneous and isotropic, with zero mean
and correlation function $\langle f({\bm
r},t)f(0,0)\rangle=\delta(t)\chi(r)$. In the simulations, the
correlation function of the forcing is taken as
$\chi(r)=\exp(-r^2/2L_f^2)$.  Contrary to the $\delta$-correlation of
the velocity, the previous hypotheses on the forcing are not
restrictive and might be easily relaxed. The anisotropy induced by the
shear makes that $C_2$ will depend both on $r$ and on the angle $\phi$
between ${\bm r}$ and the $x-$axis.  The 
steady-state solution of Eq.~(\ref{eq:7}) cannot be obtained
analytically anymore and we have to resort to numerical methods (see the
Appendix for details).

\begin{figure}
\centerline{\epsfig{file=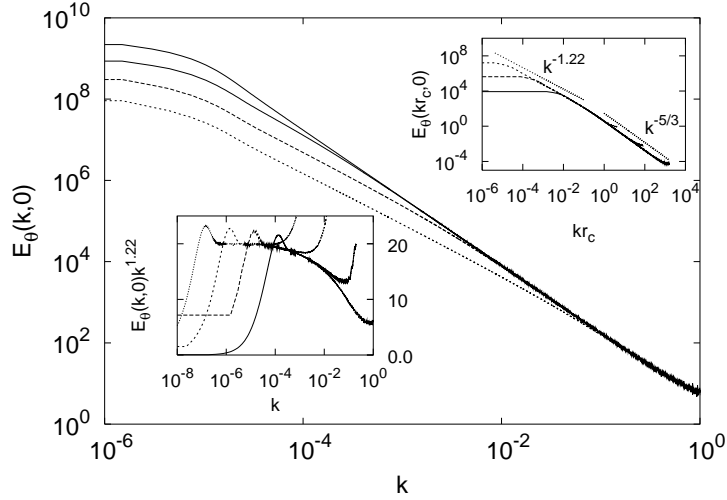,width=3.8in,angle=0}}
  \caption{Scalar spectra in the $x-$direction, $E_{\theta}(k,0)$,
obtained by fixing $D=0.05$ and varying $\sigma$ (from top to bottom,
$\sigma=0\,,\;0.0004\,,\;0.0036\,,\;0.025$). In the numerical
computations $L_{box}$ and $L_f$ have been fixed to $128^3$ and
$10^5$, respectively -- see the Appendix for the definition of the
parameters -- and the scaling exponent of the velocity $\xi=4/3$.  
The inset on the top right shows the rescaled spectra
$E_{\theta}(kr_c,0)$ {\it vs} $kr_c$. At large wavenumbers, the
$k^{-5/3}$ spectrum is recovered, while at small $k$ the power
law $k^{-11/9}$, predicted by Lagrangian dimensional arguments, 
appears.  The inset on the bottom left shows the
compensated spectra $E_{\theta}(k,0) k^{11/9}$ obtained by fixing
$\sigma=0.025$ and varying $L_f=10^4,10^5,10^6,10^7$. The number of
harmonics $\ell^{*}$, the box size $L_{box}$ and the number of grid
points have been varied accordingly.  Note that the transition region
to the asymptotic slope $-11/9$ is quite broad.}
\end{figure} 

The existence of a crossover scale, $r_c$, separating the scales
dominated by the turbulence and those affected by the shear, is easily
recognized by a direct inspection of Eq.~(\ref{eq:7}). A simple
balance of the shear and the eddy-diffusivity terms in
Eq.~(\ref{eq:7}) gives:
$r_c\sim \left(\frac{D}{\sigma}\right)^{\frac{1}{2-\xi}}$.
The crossover is well evident in Fig.~1, where the scalar spectrum for 
$\xi=4/3$ in a wide span of scales is reported. The results are in 
agreement with the Lagrangian predictions (\ref{eq:xx}),  
with $E_{\theta}(k)\sim k^{-5/3}$, for $kr_c\gg 1$, and 
$E_{\theta}(k)\sim k^{-11/9}$ asymptotically reached at the large scales. 
Moreover, as shown in the top right inset, the spectra obtained with
different shear rates all collapse by the rescaling of the wavenumbers
$kr_c$. 

The corresponding behaviour in physical space for the second-order
structure function is shown in Fig.~2a. Two regimes
are again clearly visible. A point to be
remarked in the inset is that the convergence to the asymptotic slope,
$S_2(r)\sim r^{2/9}$, is even slower than in $k$-space. The behaviour
of the second-order structure function for different orientations is
shown in Fig.~2b. As expected in a range of scales affected by the
shear, the anisotropy is strong and the crossover between the two
previous regimes depends on the angle of measurement.  It would be
very informative to have similar measurements of the anisotropy in
experimental conditions.
\begin{figure} 
\centerline{\epsfig{file=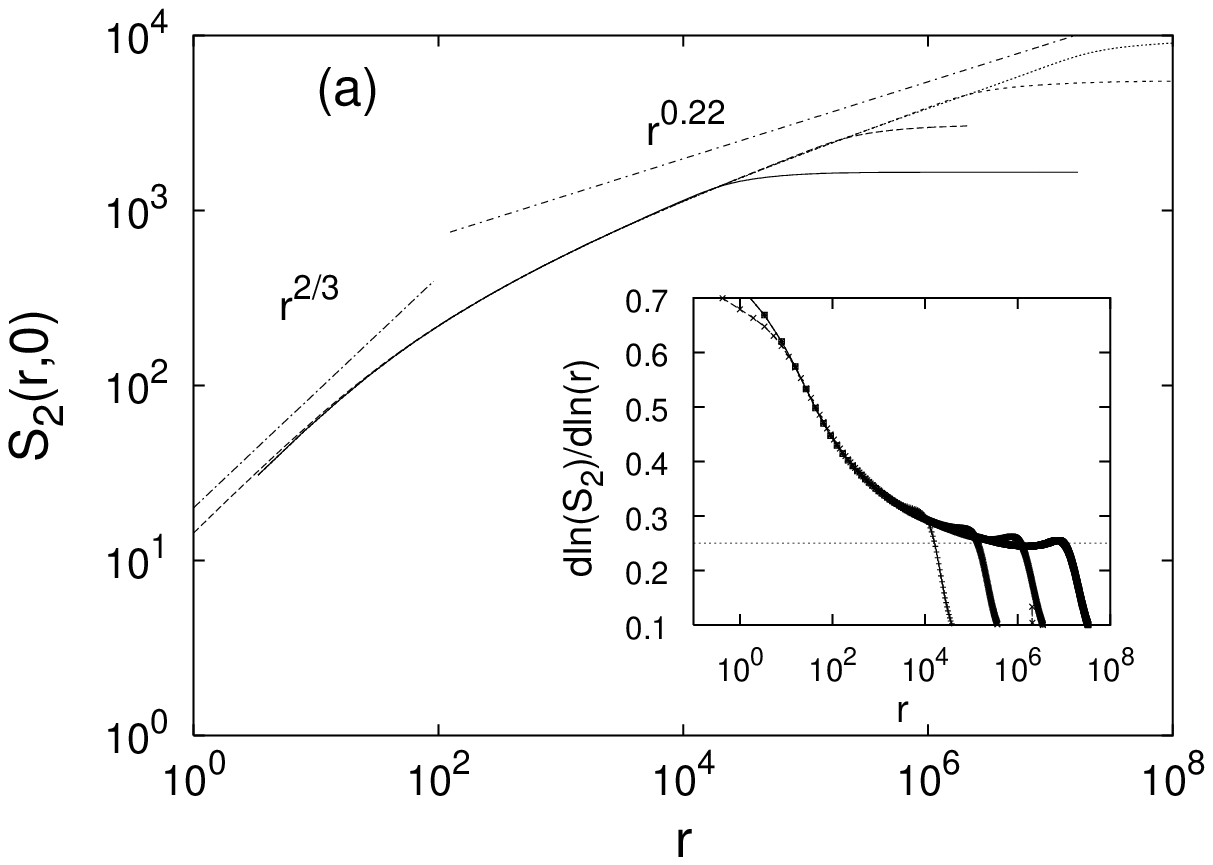,width=2.7in,angle=0}
\hfill \epsfig{file=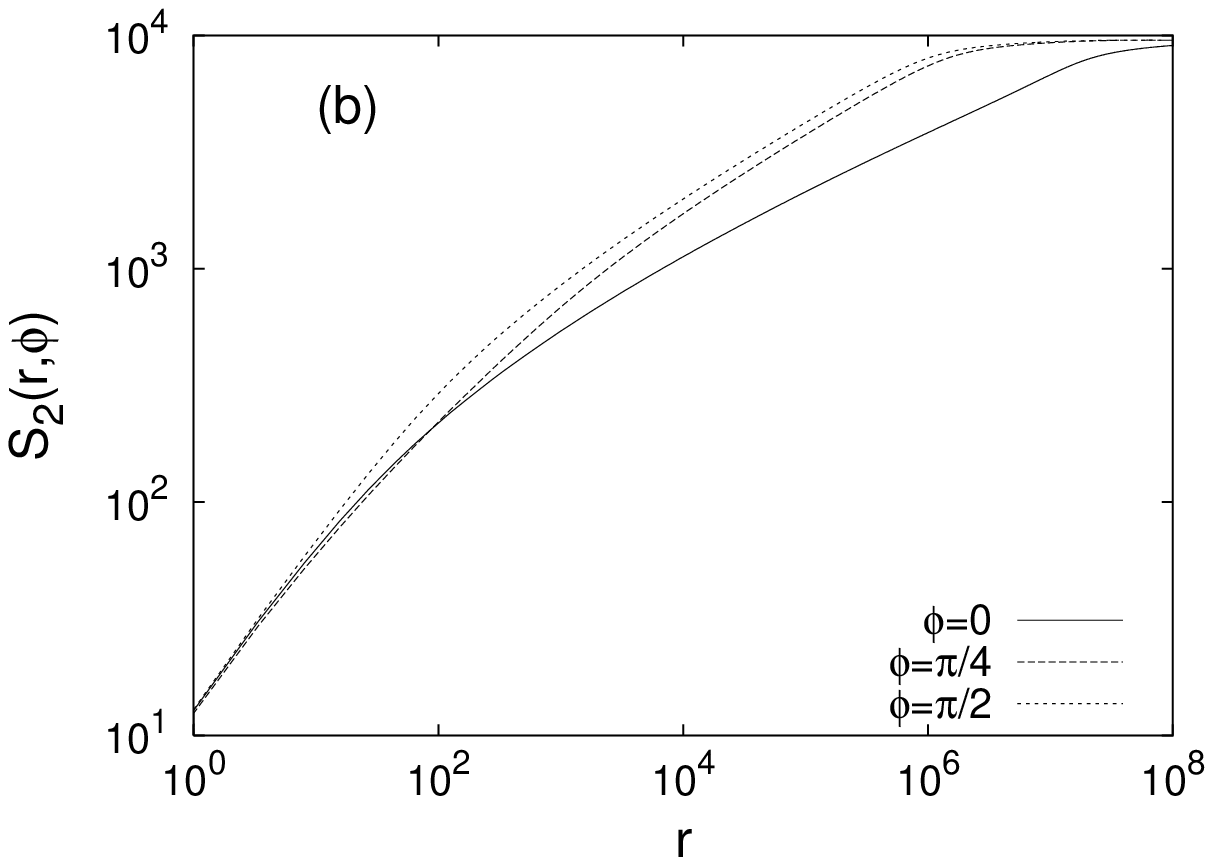,width=2.7in,angle=0}}
\caption{(a) The scalar second-order structure function measured in
the $x-$direction, $S_2(r,0)$, fixing $\sigma$ and varying $L_f$ (as
in the bottom inset of Fig.~1). The inset shows the local slopes. Note
that the approach to the asymptotic scaling $r^{0.22}$ is even slower
than in the spectra.  (b) The scalar second-order structure function
measured at different angles, $S_2(r,\phi)$, for the case
$\sigma=0.025$ and $L_f=10^7$. The statistics becomes more and more
anisotropic as larger scales are considered. A similar behaviour is
observed for the spectra (not shown).}
\end{figure} 

Note, as a side remark, that an asymptotic expansion valid
for small angles can be found analytically. Indeed, the 
correlation function $C_2$ can be expressed as
\begin{equation}
C_2(r_x,r_y)=(D/\sigma)^{-1/3} r_x^{(2-\xi)/3}
f[r_y/(D r_x^{1+\xi}/\sigma)^{1/3}]\,,
\label{piccoli1}
\end{equation}
where the angular part $f$ obeys the equation:
\begin{equation}
f''(w) + \frac{w^2}{3} f'(w) -\frac{w}{(1+\xi)} f(w) =
-\frac{\chi}{\sigma(1+\xi)}\,.
\label{piccoli2}
\end{equation}
For small angles, i.e. $w\ll 1$, the operator appearing in
(\ref{eq:7}) reduces to  $\sigma r_y \partial_{r_x} -
D(1+\xi)r_x^{\xi}\partial^2_{r_y}$ and it is easy to check that the
solution to (\ref{piccoli2}) has the form (\ref{piccoli1}) with the
function $f$ behaving as
$f(w) = -1 - \frac{\chi w^2}{2\sigma(1+\xi)} - \frac{w^3}{6(1+\xi)} + O(w^4)$.

An even closer contact to the Lagrangian arguments (\ref{eq:xx}) is made  by 
directly simulating the equations of
motion for the tracer particles.  The method used for the simulations
is the same as in \cite{Frisch98}.  Fig.~3 presents the evolution of
the particle separation for $\xi=4/3$. As long as
$\langle
R^2(t) \rangle < r_c^2\sim \left(\frac{D}{\sigma}\right)^{\frac{2}{2-\xi}}$, the Richardson prediction is
recovered and the anisotropy induced by
the shear is negligible. 
As the separation becomes larger and larger compared to $r_c$, 
the trajectories are more and more
affected by the shear and $\langle R^2\rangle \approx \langle
R_x^2\rangle \gg \langle R_y^2\rangle$. In this range of scales,
$\langle R_x^2\rangle$ and $\langle R_y^2\rangle$ grow with two
different power laws and the local slopes in the inset of 
Fig.~3 are found to be in agreement with (\ref{eq:xx}).
The
predictions (\ref{eq:xx}) and the 
numerical simulations
are found to be in agreement for values of $\xi\neq 4/3$ as well (data not shown). 

An interesting remark is that
the Lagrangian numerics evidences a wide crossover region between
the two different regimes (see Fig.~3). That is the
reason for the very slow convergence of the
scalar structure functions and spectra to their asymptotic slopes. 
In general, detecting the
transition between the inertial-convective and the shear regimes demands
a very broad range of accessible scales.

\begin{figure} 
\centerline{\epsfig{file=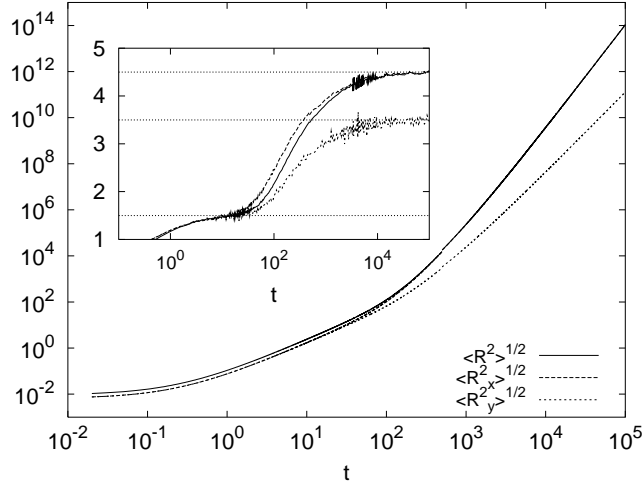,width=3.8in,angle=0}}
\caption{The curves of the Lagrangian separations 
as a function of time, for a Kraichnan flow with exponent $\xi=4/3$. 
At small times, $\langle R_x^2(t)\rangle=\langle
R_y^2(t)\rangle=\langle R^2(t)\rangle/2 \sim t^{3}$, as expected
according to the Richardson law. At large times, $\langle
R^2(t)\rangle\approx\langle R_x^2(t)\rangle\sim t^{9}$ and $\langle
R_y^2(t)\rangle\sim t^{7}$, in agreement with (\ref{eq:xx}). 
The inset displays the local slopes as a
function of time. Note the wide crossover region between the two
regimes. The numerical integrations of Eqs.~(\ref{eq:12a}) and
(\ref{eq:12b}) have been performed following the point-splitting
scheme described in \protect\cite{Frisch98}, i.e.  with molecular
diffusivity $\kappa=0$ and taking the initial separation small but
finite. }
\end{figure} 

We finally consider the case where the shear is random. If its
direction is fixed and the time-scale of its intensity 
is short, the resulting exponent
of the scalar spectrum is $-2+\xi/2$. If both the direction and the
intensity of the shear rapidly fluctuate in time, i.e. the shear is a
Kraichnan flow with $\xi=2$, isotropy is recovered and the scalar
correlation function in the stationary state obeys
$(1/r)\partial_r (\sigma r^3+ D
r^{1+\xi}+ 2\kappa r)\partial_r C_2 =-\chi(r)$.
The equation is easily solved analytically and the ensuing slopes of
the scalar spectra are $-3+\xi$ at small scales (large wavenumbers)
and $-1$ at large scales (small wavenumbers).

\section{Conclusions}
The results presented here illustrate the presence of a new shear
regime for a passive scalar transported in turbulent shear flows.  The
scale of crossover between the classical O\-bukh\-ov-Corrsin range and the
larger scales, where the effects of the shear are felt, is 
proportional to the ratio between the turbulent and the shear intensities. 
The range of scales where the
shear-regime is observed becomes then parametrically large with increasing the
shear rate.  In experiments where the latter is fixed, increasing the
turbulence intensity has the double effect of decreasing the
Kolmogorov scale and increasing the crossover length $r_c$. The
Obukhov-Corrsin scaling is therefore observed over a broader and
broader range of scales. When $r_c$ is much larger than the integral
scale $L$ of the scalar, the picture at scales $\leq L$ becomes
essentially shear-free.  For $r_c\simeq L$, the presence of the
shear-range is still relevant due to the crossover effects which
induce a scalar spectrum shallower than $-5/3$ and anisotropic.
Those effects disappear quite slowly with the Reynolds number due to
the broad span of scales required for the crossover. Furthermore, we
have shown that the scalar spectrum in the shear-region is sensitive 
to the geometry and the fluctuations of the shear.
Experiments with different set-up's, where the shear components 
have drastically differing properties, might then yield different scalings 
at asymptotically large scales. All are shallower than $-5/3$, 
though, and therefore inducing crossover effects that tend to 
flatten out the scalar spectrum. 
In those conditions where shear fluctuations in intensity 
and direction are kept controlled, we specifically predict 
the existence of a $k^{-4/3}$ 
asymptotic regime. 
The previous picture seems to account for the major points in the
experimental observations presented in~\S\,2.  

Two issues for future
investigation are the following. Theoretically, it is quite likely
that the shear will affect the scalar higher-order statistics;
the resulting behaviours remain to be clarified.  Experimentally, it
would be of major interest to have a set-up where the Reynolds number
is kept fixed and the shear rate is increased, in well-controlled conditions, 
so as to enlarge the shear-regime and quantitatively analyze its properties.

\begin{acknowledgments}
The authors acknowledge the hospitality of the Erwin Schr\"odinger
Institute in Vienna.  This work has been partially supported by the EU
under the contract ``Non-Ideal Turbulence'' HPRN-CT-2000-00162 and by 
Cofin 2001 (prot. 2001023848).
\end{acknowledgments}

\appendix
\section{Equations for the correlation function and numerical computation}
The anisotropy introduced by the shear can be handled by using polar
coordinates, which allow to decompose the correlation function as:
\begin{equation}
C_2({\bm r},t)=C_2(r,\phi,t)= F_0(r,t)+\sum_{\ell=1}^{\infty} F_\ell(r,t)
\cos(2 \ell \phi) +\sum_{\ell=1}^{\infty} G_\ell(r,t) \sin(2 \ell
\phi) \,.
\label{eq:8}
\end{equation}
Here, $\ell$ labels the order of the harmonics. Note that only even
harmonics are present for symmetry reasons.  Substituting (\ref{eq:8})
into (\ref{eq:7}), one obtains the following set of 1+1 dimensional
partial differential equations:
\begin{eqnarray}
\partial_t F_{\ell} &&+ \sigma \frac{r}{4} [\partial_r G_{\ell + 1} -
  \partial_r G_{\ell - 1}] + \frac{\sigma}{2} [(\ell + 1)G_{\ell+1} -
  2\ell G_{\ell} + (\ell - 1)G_{\ell-1}] = \nonumber \\ &&\frac{1}{r}
\partial_r [2\kappa r+D r^{1+\xi}] \partial_r F_{\ell}
-[2\kappa+(1+\xi) D r^{\xi}] \frac{4 \ell^2}{r^2} F_{\ell}
+\delta_{\ell,0} \chi \,,
\label{eq:ap1} \\
\partial_t G_{\ell} &&- \sigma \frac{r}{4} [\partial_r F_{\ell + 1}
 -   (1+\delta_{\ell,1}) \partial_r F_{\ell-1}] -\frac{\sigma}{2}
[(\ell + 1)F_{\ell+1} - 2\ell F_{\ell} + (\ell - 1)F_{\ell-1}] = \nonumber \\
&& \frac{1}{r} \partial_r
\left[2\kappa r+D r^{1+\xi} \right]\partial_r
G_{\ell}-[2\kappa+(1+\xi)D r^{\xi}] \frac{4 \ell^2}{r^2}G_{\ell}\,. \label{eq:ap2} 
\end{eqnarray}
Note that the coupling among the harmonics is indeed
due to the shear term.

Those equations have been numerically integrated for times long enough
to reach a stationary state. The correlation function $C_2(r,\phi)$
is then computed at different orientations $\phi$ from $F_{\ell}(r)$
and $G_{\ell}(r)$ by using (\ref{eq:8}). The second-order structure
function $S_2(r,\phi)=2(C_2(0)-C_2(r,\phi))$ easily follows, as well
as the scalar spectrum $E_{\theta}(k,\phi)$, which is obtained by a
one-dimensional Fourier transform.

The equations have been solved for various choices of the
parameters. In practice, we have fixed $D$ and varied $\sigma$ and $L_f$.
As for the boundary conditions at the origin, we have taken $F'_0(0)=0$,
$G_{\ell}(0)=0$ for each $\ell$ and $F_{\ell}(0)=0$ for ${\ell}>0$.
The infrared boundary condition has been fixed by setting to zero the values
of all the functions $F_{\ell}(r)$ and $G_{\ell}(r)$ beyond
$r=L_{box}$. The maximum size of the system, $L_{box}$, has been taken much
larger than the forcing scale $L_f$ to properly resolve the
anisotropic contributions.


\begin{thebibliography}{} 

\bibitem[Batchelor (1959)]{Batchelor59}
{\sc Batchelor, G.K.} 1959
{Small-scale variation of convected quantities like temperature in turbulent
fluid. Part 1. General discussion and the case of small conductivity.}
{\it J. Fluid Mech.}~{\bf 5}, 113--133.

\bibitem[Corrsin (1951)]{Corrsin51}
{\sc Corrsin, S.} 1951
{On the spectrum of isotropic temperature field in isotropic turbulence.}
{\it J. Appl. Phys.} {\bf 22}, 469--473.

\bibitem[Falkovich, Gaw\c{e}dzki \& Vergassola (2001)]{Falkovich01}
{\sc Falkovich, G., Gaw\c{e}dzki, K. \& Vergassola, M.} 2001
{Particles and fields in fluid turbulence.}
{\it Rev. Mod. Phys.} {\bf 73}, 913--975.

\bibitem[Frisch et al. (1998)]{Frisch98} {\sc Frisch, U., Mazzino,
A. \& Vergassola, M.}  1998 {Intermittency in passive scalar advection}  
{\it Phys. Rev. Lett.} {\bf 80}, 5532--5535.

\bibitem[Jayesh, Tong \& Warhaft (1994)]{Jayesh94}
{\sc Jayesh, Tong, C. \& Warhaft, Z.} 1994
{On temperature spectra in grid turbulence.}
 {\it Phys. Fluids} {\bf 6}, 306--312.

\bibitem[Kraichnan (1968)]{Kraichnan68} {\sc Kraichnan, R. H.} 1968
{Small-scale structure of a scalar field convected by turbulence.}
{\it Phys. Fluids} {\bf 11}, 945--963.

\bibitem[Levèque (1928)]{Leveque28}
{\sc Levèque, M., A.} 1928
{Les lois de la transmission de la chaleur par convection.}
{\it Ann. Mines} {\bf 13}, 201-239.

\bibitem[Mestayer (1982)]{Mestayer82}
{\sc Mestayer, P.} 1982
{Local isotropy and anisotropy in a high-Reynolds
    number turbulent boundary layer.}
{\it J. Fluid Mech.} {\bf 125}, 475-503.

\bibitem[Miller \& Dimotakis (1996)]{Miller96}
{\sc Miller, P., L. \& Dimotakis, P., E.} 1996
{Measurement of scalar power spectra in high Schmidt number turbulent jets.}
{\it  J. Fluid Mech.} {\bf 308}, 129--146.

\bibitem[Mydlarski \& Warhaft (1998)]{Mydlarski98}
{\sc Mydlarski, L. \& Warhaft, Z.} 1998
{Passive scalar statistics in  high-P\'eclet-number grid turbulence.}
{\it  J. Fluid Mech.} {\bf 358}, 135--175.

\bibitem[Obukhov (1949)]{Obukhov49}
{\sc Obukhov, A. M.} 1949
{The structure of the temperature field in a turbulent flow}
{\it Izv. Akad. Nauk. SSSR, Ser. Geogr. and Geophys.} {\bf 13}, 58--69.

\bibitem[Siggia (1994)]{Siggia94}
{\sc Siggia, E. D.} 1994 
{High Rayleigh number convection}
{\it Annu. Rev. Fluid Mech.} {\bf 26}, 137-168.

\bibitem[Sreenivasan (1991)]{Sreenivasan91} 
{\sc Sreenivasan, K. R.} 1991 
{On local isotropy of passive scalars in turbulent
    shear flows.} 
 {\it Proc. R. Soc. London Ser. A} {\bf 434}, 165--182. 

\bibitem[Sreenivasan (1996)]{Sreenivasan96} 
{\sc Sreenivasan, K. R.} 1996 
{The passive scalar spectrum and the Obukhov-Corrsin
    constant.} 
 {\it Phys. Fluids} {\bf 8}, 189--196. 

\bibitem[Staicu \& van de Water (2003)]{Staicu03}
{\sc Staicu, A. \& van de Water, W.} 2003
{Small-scale velocity jumps in shear turbulence.}
{\it Phys. Rev. Lett.} {\bf 90}, 094501-094505.

\bibitem[Villermaux, Innocenti \& Duplat (2001)]{Villermaux01} 
{\sc Villermaux, E.,  Innocenti, C. \&  Duplat, J.} 2001
{Short circuits in the Corrsin-Obukhov cascade.}
{\it Phys. Fluids} {\bf 13}, 284--289.

\bibitem[Warhaft (2000)]{Warhaft00}
{\sc Warhaft, Z.} 2000
{Passive scalars in turbulent flows.}
{\it Ann. Rev. Fluids Mech.}  {\bf 32}, 203--240. 

\bibitem[Yeung, Xu \& Sreenivasan (2002)]{Yeung02} 
{\sc Yeung, P.K., Xu, S. \& Sreenivasan, K.R.} 2002
{Schmidt number effects on turbulent transport with uniform mean scalar gradient}
{\it Phys. Fluids} {\bf 14}, 4178--4191.

\end{thebibliography}
\end{document}